\documentclass[twocolumn,preprintnumbers,amsmath,amssymb,showpacs]{revtex4}
\usepackage{epsfig}
\usepackage{color}
\begin{document}
\setlength{\voffset}{1.0cm}
\title{Covariant boost and structure functions of baryons in Gross-Neveu models}
\author{Wieland Brendel\footnote{wieland@theorie3.physik.uni-erlangen.de}}
\author{Michael Thies\footnote{thies@theorie3.physik.uni-erlangen.de}}
\affiliation{Institut f\"ur Theoretische Physik III,
Universit\"at Erlangen-N\"urnberg, D-91058 Erlangen, Germany}
\date{\today}
\begin{abstract}
Baryons in the large $N$ limit of two-dimensional Gross-Neveu models are reconsidered. The time-dependent Dirac-Hartree-Fock approach 
is used to boost a baryon to any inertial frame and shown to yield the covariant energy-momentum relation. Momentum distributions
are computed exactly in arbitrary frames and used to interpolate between the rest frame and the infinite momentum frame, where they are
related to structure functions. Effects from the Dirac sea depend sensitively on the occupation fraction of the valence level and the bare fermion
mass and do not vanish at infinite momentum. In the case of the kink baryon, they even lead to divergent quark and antiquark structure functions
at $x=0$.
\end{abstract}
\pacs{11.10.Kk,11.10.St,11.30.Cp}
\maketitle

\section{Introduction}\label{sect1}

Given the notorious difficulty of solving non-perturbative problems in quantum field theories (QFT) systematically, it is of theoretical interest to study
exactly soluble model field theories. From the point of view of strong interaction physics, a particularly gratifying example is the Gross-Neveu model
family in 1+1 dimensions with Lagrangian \cite{1}
\begin{equation}
{\cal L} = \bar{\psi} \left( {\rm i} \partial \!\!\!/ -m_0\right)\psi + \frac{g^2}{2} \left[ (\bar{\psi}\psi)^2+ \lambda (\bar{\psi} {\rm i} \gamma_5 \psi)^2\right].
\label{A1}
\end{equation}
It describes $N$ species of self-interacting fermions (flavor labels are suppressed as usual) with discrete
($\lambda=0$, Gross-Neveu (GN) model) or continuous ($\lambda=1$, 
Nambu--Jona-Lasinio (NJL) model \cite{2}) chiral symmetry, possibly broken by the bare mass term $\sim m_0$. Throughout this work we are only interested
in the 't~Hooft limit $N\to \infty, Ng^2 = {\rm const.}$ 

The simple Lagrangian (\ref{A1}) gives rise to a number of interesting phenomena such as asymptotic freedom,
spontaneous symmetry breaking, hadronic bound states and non-trivial phase diagrams as a function of temperature and chemical potential.
These non-perturbative phenomena are either accessible through analytical methods, or else numerically to any desired
accuracy. So far, the main focus has been on static properties such as the hadron spectrum \cite{3,4,5,6,7,8,9,10,11} or equilibrium thermodynamics 
\cite{12,13,14,15,16}. We see no reason why such models should not be instructive for dynamical problems as well. 
In principle, it is known how to generalize the semiclassical methods adequate in the large $N$ limit 
to the time-dependent case \cite{17,18,3}. For baryons in particular, this amounts to replace the relativistic Hartree-Fock (HF)
approach by its time-dependent generalization. The non-relativistic version of time dependent Hartree-Fock (TDHF), introduced originally by Dirac \cite{19},
is well-known in nuclear and heavy-ion physics \cite{20}. This paper is a first step towards attacking time dependent problems in QFT with
relativistic many-body techniques. We hope to
extend thereby the spectrum of questions which can be solved exactly in field theory models, serving as a testing ground for other approaches.

Consider a baryon in the Gross-Neveu model, i.e., a HF state of $n$ valence quarks with the polarized Dirac sea.
Dashen, Hasslacher and Neveu (DHN) have already addressed the issue of covariance of such bound states formally from the point of view of semiclassical
 methods and concluded
that the relativistic energy-momentum relation $E=\sqrt{M^2+P^2}$ is satisfied \cite{17}. Here we take up this question using
relativistic many-body methods. Assuming that the baryon state has been constructed in the rest frame, we boost it to an arbitrary inertial frame
and obtain a self-consistent solution of the Dirac-TDHF equations. We then compute energy and momentum of the moving baryon and verify 
covariance in the large $N$ limit. In general, to compute the energy and the momentum 
of such bound states in QFT is rather delicate. It requires renormalization, vacuum subtraction and the use of a non-covariant cutoff and a finite box
at intermediate stages, at least in the previous approaches in the rest frame \cite{3,4,5,10}. In view of these complications, we find it worthwhile to
examine in detail how covariance is maintained, even if the outcome is 
known on general grounds. As a matter of fact, we will not need the explicit solution of the baryon in the rest frame (which is known analytically
only for the GN model, but not for the NJL model). This sheds some light on how covariance emerges in the canonical formalism
where it is often deeply hidden.

Having confirmed covariance, as an application we compute quark momentum distributions for the baryon
moving with arbitrary velocity. Here we do need the explicit HF wave functions, so that we can carry out an analytical calculation for the DHN baryons 
of the GN model only. By taking the infinite momentum frame limit, we also determine structure functions, the quantities of central interest in deep
inelastic scattering and quantum chromodynamics (QCD), see e.g. \cite{21}. In such a model study it is easy to disentangle
valence from sea quarks and to study explicitly the dependence of the structure functions on the occupation of the valence level (or baryon number)
and quark masses.
  
We will also address the question to which extent the HF calculation simplifies in the infinite momentum frame, relevant for
attempts to work in light-cone quantization \cite{22,23}. In the meson problem (quark-antiquark bound state), such simplifications do indeed occur and 
have been instrumental for the first successful application of light-cone quantization by 't Hooft to large $N$ QCD$_2$ \cite{24}. Here as well as in the GN model,
equivalence of the meson spectrum with equal time quantization 
has been firmly established long ago \cite{25,26,27,28}. In the case of baryons, the situation is less clear. We are aware of a few early attempts
to compute baryon masses
on the light-cone for finite $N$ QCD$_2$ \cite{29,30} and of the works \cite{31,32,33} relevant for the large $N$ limit, but there has been no evaluation
of the DHN baryons on the light-cone
to the best of our knowledge. In the meson case, a pure valence quark approach becomes exact in light-cone quantization, technically reducing the
relativistic random phase approximation to the simpler Tamm-Dancoff approximation.
By contrast, the role of the Dirac sea for baryons in the infinite momentum frame is still unclear. 

This paper is organized as follows: In Sec.~\ref{sect2}, we briefly recall the evaluation of the baryon mass in the rest frame and show how
this calculation can be simplified significantly. This explains several cancellations which were observed in earlier works and yields
a new and surprisingly simple relationship between the baryon mass and the self-consistent HF potential. This step is necessary to keep the
proof of covariance in later sections tractable. In Sect.~\ref{sect3} we set up the Dirac-TDHF approach for boosting the baryon. Sect.~\ref{sect4} is dedicated to the 
calculation of baryon energy and momentum in an arbitrary inertial frame and to the test of covariance. In Sect.~\ref{sect5} we define and compute  
quark and antiquark momentum distributions for the moving DHN baryon of the GN model with (broken or unbroken) discrete chiral symmetry.
Upon performing the limit to the infinite momentum frame, we arrive at simple expressions for structure functions in Sect.~\ref{sect6}. 
We also discuss issues like
valence and sea quark contributions, or the dependence on the baryon number and the bare fermion mass. We finish this paper with a concluding
section, Sec.~\ref{sect7}.
 
\section{Simplifying the calculation of the baryon mass in the rest frame}\label{sect2}

The calculation of the baryon mass in the GN model \cite{3,5} or of the Shei bound state in the NJL model \cite{4,10}
is rather involved. The corresponding calculation of energy and momentum of the moving baryon
is even more difficult. If one looks at the HF calculation in the rest frame in detail, one finds 
that many terms cancel at the end. This points to the presence of simplifying
features. In order to take advantage of these simplifications in the case of the moving baryon, we first have to
gain a better understanding of the calculation in the rest frame. This is the topic of the present section.

Assume that we have solved the HF problem self-consistently for given baryon number,
\begin{equation}
\left[ -{\rm i}\gamma_5 \partial_x + \gamma^0 S(x) + {\rm i} \gamma^1 \lambda P(x) \right] \phi_{\alpha}(x)  =  E_{\alpha} \phi_{\alpha}(x),
\label{E1}
\end{equation}
with scalar and pseudo-scalar potentials
\begin{eqnarray}
S(x) & = & - g^2 \sum_{\alpha}^{\rm occ} \bar{\phi}_{\alpha}(x)\phi_{\alpha}(x) +m_0,
\nonumber \\
P(x) & = & - g^2 \sum_{\alpha}^{\rm occ} \bar{\phi}_{\alpha}(x){\rm i}\gamma_5 \phi_{\alpha}(x). 
\label{E2}
\end{eqnarray}
The sums run over all occupied states, i.e., the valence level and the Dirac sea.
The Hamiltonian and momentum operator of the GN model read
\begin{eqnarray}
H &=& \int {\rm d}x \left\{ -{\rm i} \psi^{\dagger}\gamma_5 \partial_x \psi + m_0 \bar{\psi}\psi - \frac{g^2}{2} \left[ (\bar{\psi}\psi)^2 
\right. \right.
\nonumber \\
& & \left. \left. + \lambda (\bar{\psi}{\rm i}\gamma_5 \psi)^2\right] \right\},
\nonumber \\
P & = & \int {\rm d}x \left( -{\rm i}\psi^{\dagger} \partial_x \psi \right).
\label{E3}
\end{eqnarray}
Let us write the expectation values in the HF baryon state as
\begin{eqnarray}
\langle H \rangle   & = & \int {\rm d}x \langle{\cal H}(x)\rangle,
\nonumber \\
\langle P \rangle & = & \int {\rm d}x \langle {\cal P}(x) \rangle.
\label{E4}
\end{eqnarray}
where the energy and momentum densities are given by
\begin{eqnarray}
\langle {\cal H} \rangle
& = & \sum_{\alpha}^{\rm occ}\left\{ -{\rm i}\phi_{\alpha}^{\dagger} \gamma_5 \partial_x \phi_{\alpha} + \frac{1}{2}(m_0+S)
\bar{\phi}_{\alpha} \phi_{\alpha} 
\right.
\nonumber \\
& & \left. + \frac{1}{2} \lambda P \bar{\phi}_{\alpha}{\rm i} \gamma_5 \phi_{\alpha} \right\},
\nonumber \\
\langle {\cal P} \rangle
& = & \sum_{\alpha}^{\rm occ} \left(-{\rm i} \phi_{\alpha}^{\dagger} \partial_x \phi_{\alpha} \right). 
\label{E5}
\end{eqnarray}
We have to subtract the vacuum contribution from $\langle {\cal H} \rangle$. For ease of notation, we denote the expectation values
$\langle {\cal H} \rangle,\langle {\cal P} \rangle$ generically
by ${\cal F}$ and their contribution from single particle state $\alpha$ by ${\cal F}^{\alpha}$,
\begin{equation}
{\cal F} =  \sum_{\alpha}^{\rm occ} {\cal F}^{\alpha}.
\label{E6}
\end{equation}
The $c$-number densities ${\cal F}^{\alpha}$ are $x$ dependent, approaching a constant
${\cal F}_{\rm asy}^{\alpha}$ for $x\to \pm \infty$.
After splitting off the constant term,
\begin{equation}
{\cal F}^{\alpha} = ({\cal F}^{\alpha}-{\cal F}_{\rm asy}^{\alpha}) + {\cal F}_{\rm asy}^{\alpha}
:= {\cal F}_{\rm loc}^{\alpha} +  {\cal F}_{\rm asy}^{\alpha},
\label{E7}
\end{equation} 
the localized part ${\cal F}_{\rm loc}^{\alpha}$ can be integrated over $x$ from $- \infty$ to $+ \infty$ with a finite
result. The constant, asymptotic part ${\cal F}_{\rm asy}$ requires the usual careful vacuum subtraction. To this end we enclose the system
temporarily in a box of length $L$ and impose periodic boundary conditions.  
${\cal F}_{\rm asy}^{\alpha}$ is non-vanishing only for the continuum states and differs from the corresponding vacuum quantity
due to a change in the density of states,
\begin{equation}
{\cal F}_{\rm asy}^{\alpha} = {\cal F}_{\rm asy}(k) = {\cal F}_{\rm vac}(k) \left( 1 - \frac{1}{L} \frac{{\rm d}\delta(k)}{{\rm d}k} \right).
\label{E8}
\end{equation}
Here, $k$ labels the asymptotic momentum of the scattering state $\alpha$ and $\delta(k)$ is the scattering phase shift.
Owing to the finite box, the fermion momenta are discretized as $k_n^0=2\pi n/L$ in the vacuum and as
\begin{equation}
k_n = k_n^0 - \frac{1}{L}\delta(k_n)
\label{E9}
\end{equation}
in the baryon state. Subtracting the vacuum contribution, we evaluate
\begin{eqnarray}
& & \sum_n \left( {\cal F}_{\rm asy}(k_n) -  {\cal F}_{\rm vac}(k_n^0)\right) 
\nonumber \\
&  \approx  & \sum_n \left({\cal F}_{\rm asy}(k_n^0) - \frac{1}{L} \delta(k_n^0)
 \left. \frac{{\rm d}{\cal F}_{\rm asy}(k)}{{\rm d} k}\right|_{k_n^0}-{\cal F}_{\rm vac}(k_n^0) \right)
\nonumber \\
& \approx & - \frac{1}{L} \sum_n \left(  \delta(k_n^0)
 \left. \frac{{\rm d}{\cal F}_{\rm vac}(k)}{{\rm d} k}\right|_{k_n^0} +   \left. \frac{{\rm d}\delta(k)}{{\rm d}k}\right|_{k_n^0} {\cal F}_{\rm vac}(k_n^0) \right)
\nonumber \\
& \approx & - \int \frac{{\rm d}k}{2\pi} \frac{{\rm d}}{{\rm d}k} \left[ \delta(k){\cal F}_{\rm vac}(k) \right]
\nonumber \\
& = & -  \frac{1}{2\pi} \lim_{k \to \infty} \left[ \delta(k) {\cal F}_{\rm vac}(k) - \delta(-k) {\cal F}_{\rm vac}(-k) \right]
\label{E10}
\end{eqnarray}
where we have used Eq.~(\ref{E8}) and taken the limit $L\to \infty$. Hence the vacuum subtracted asymptotic part is a pure surface term
in momentum space.
The vacuum energy and momentum densities are 
\begin{eqnarray}
L\langle {\cal H}_{\rm vac}(k)\rangle &=& - \frac{2k^2+m^2+mm_0}{2 E_k}N,
\nonumber \\
L\langle {\cal P}_{\rm vac}(k)\rangle & = & kN.
\label{E11}
\end{eqnarray}
Anticipating that $\delta(k) \sim 1/k$ for large $|k|$ (as will be confirmed in the eikonal approximation below), the asymptotic, $x$-independent
contribution to the baryon energy and momentum in the rest frame finally becomes
\begin{eqnarray}
L\sum_n \left[ \langle {\cal H}_{\rm asy}(k_n) \rangle -\langle {\cal H}_{\rm vac}(k_n^0)\rangle \right] & = & \frac{N}{\pi} \lim_{k \to \infty} k \delta(k),
\nonumber \\
L\sum_n \left[ \langle {\cal P}_{\rm asy}(k_n) \rangle -\langle {\cal P}_{\rm vac}(k_n^0)\rangle \right] & = & 0.
\label{E12}
\end{eqnarray}
We now turn to the localized part. Here it is useful to go back to the energy momentum tensor, since its local
conservation law leads to drastic simplifications in 1+1 dimensions.
The canonical energy momentum tensor reads \cite{34}
\begin{equation}
{\cal T}_{\mu \nu}  =  {\rm i} \bar{\psi} \gamma_{\mu} \partial_{\nu} \psi - g_{\mu \nu} {\cal L}
\label{E13} 
\end{equation}
with the Lagrangian density ${\cal L}$ from Eq.~(\ref{A1}).
Working out its components, we find
\begin{eqnarray}
{\cal T}_{00} & = & {\cal H} 
\nonumber \\
& =& -{\rm i} \psi^{\dagger} \gamma_5 \partial_x \psi + m_0 \bar{\psi}\psi - \frac{g^2}{2}\left[ (\bar{\psi}\psi)^2
+\lambda (\bar{\psi}{\rm i}\gamma_5 \psi)^2\right],
\nonumber \\
{\cal T}_{11} & = & {\cal H} -m_0 \bar{\psi}\psi,
\nonumber \\
{\cal T}_{01} & = & {\rm i} \psi^{\dagger} \partial_x \psi \ = \ - {\cal P},
\nonumber \\
{\cal T}_{10} & = & - {\cal P} - \frac{\rm i}{2} \partial_{\mu}j_5^{\mu},
\label{E14}
\end{eqnarray}
where the divergence of the axial current $j_5^{\mu} = \bar{\psi} \gamma^{\mu} \gamma_5 \psi$ is given by
\begin{equation}
\partial_{\mu}j_5^{\mu} = 2(m_0 -  (1-\lambda) g^2  \bar{\psi}\psi) \bar{\psi}{\rm i}\gamma_5 \psi.
\label{E15}
\end{equation}
For expectation values in any stationary state, the conservation law
\begin{equation}
\partial^{\mu} {\cal T}_{\mu \nu}  =  0
\label{E16}
\end{equation}
reduces  to
\begin{eqnarray}
\partial_x \left( {\cal H} - m_0 \bar{\psi}\psi \right) & = & 0 ,
\label{E17} \\
\partial_x \left( {\cal P}  + \frac{\rm i}{2} \partial_x \rho \right) & = & 0.
\label{E18}
\end{eqnarray}
Here, we have made use of the elementary fact that ($j^{\mu}=\bar{\psi}\gamma^{\mu}\psi$)
\begin{equation}
j_5^0 = j^1, \qquad j_5^1 = j^0 = \rho
\label{E19}
\end{equation}
in 1+1 dimensions. 
For the local pieces, we therefore get
\begin{eqnarray}
\langle {\cal H}_{\rm loc} \rangle & = & m_0 \langle \bar{\psi}\psi \rangle_{\rm loc},
\nonumber \\
\langle {\cal P}_{\rm loc} \rangle  & = & - \frac{\rm i}{2} \partial_x \rho.
\label{E20}
\end{eqnarray}
The first equation implies in particular that the energy density in the chiral limit of the GN model should become $x$-independent. This is 
responsible for cancellations encountered if one evaluates all terms separately, see the GN model \cite{3,5} and the Shei bound state \cite{4,10}.
Integrating the densities (\ref{E20}) over $x$ and adding the $x$-independent, vacuum subtracted contributions 
(\ref{E12}), we arrive at a simple expression for the baryon mass 
\begin{equation}
M_{\rm B}  =  \frac{N}{\pi} \lim_{k \to \infty} k \delta(k) +  m_0 \int_{-\infty}^{\infty} {\rm d}x \langle \bar{\psi}\psi \rangle_{\rm loc},
\label{E21}
\end{equation}
and trivially $P_{\rm B}=0$ in the rest frame. It is now easy to express the right-hand-side of Eq.~(\ref{E21})  in terms of the HF potential.
The asymptotic behavior of the phase shift can be determined with the help of the eikonal approximation (Glauber theory \cite{35}).
To this end, we write down the stationary Dirac equation, eliminate the lower spinor component $v$ and use
the following ansatz for the upper component,
\begin{equation}
u(x) = \tilde{u}(x) {\rm e}^{{\rm i}kx},
\label{E22}
\end{equation}
with slowly varying modulation $\tilde{u}(x)$. In the high energy limit one then finds a first order differential 
equation
\begin{equation}
\frac{{\rm d}\tilde{u}}{{\rm d}x} = \frac{\rm i}{2k} \left( S' +{\rm i}\lambda P' -S^2-\lambda P^2 + m^2 \right) \tilde{u}
\label{E23}
\end{equation}
from which the asymptotic scattering phase shift for $|k|\to \infty$ can be deduced as
\begin{equation}
\delta(k) = - \frac{1}{2k} \int_{-\infty}^{\infty} {\rm d}x \left( S^2+ \lambda P^2 -m^2 \right)
\label{E24}
\end{equation}
(assuming no surface term from the derivatives).
The chiral condensate on the other hand is directly related to the HF potential by self-consistency,
\begin{equation}
 m_0 \int_{-\infty}^{\infty} {\rm d}x \langle \bar{\psi}\psi \rangle_{\rm loc} = - \frac{m\gamma N }{\pi} \int_{- \infty}^{\infty} {\rm d}x (S-m).
\label{E25}
\end{equation}
We have introduced the confinement parameter \cite{36}
\begin{equation}
\gamma = \frac{\pi}{Ng^2}\frac{m_0}{m}.
\label{E26}
\end{equation}
The final relation between the baryon mass and the self consistent potential is then
\begin{equation}
M_{\rm B} = - \frac{N}{2\pi} \int_{-\infty}^{\infty} {\rm d}x \left[ (S+ \gamma m)^2 + \lambda P^2 - (m + \gamma m)^2 \right].
\label{E27}
\end{equation}
This formula holds in massless and massive GN or NJL models. If we insert the known potential $S$ for the massive GN model for instance, we recover
the result given in Eq.~(\ref{H32}) below.
We have also checked the formula for the baryon of the massive NJL model
in the derivative expansion \cite{8} and found agreement with known results up to O($m_{\pi}^{11}$). In the chiral limit of the NJL model ($\lambda=1, \gamma=0$),
one sees nicely the appearance of a massless baryon \cite{37} if the potential traces out the chiral circle, $S^2+P^2=m^2$.

Concluding this section, we note that the cancellations observed in previous calculations of the baryon mass have two distinct sources: The fact that the
vacuum subtracted, constant part of the energy density is a pure surface term in momentum space, and local energy-momentum conservation  
relating the $x$-dependent part of the energy density to the subtracted chiral condensate.
These observations will help us to evaluate energy and momentum of the moving baryon more efficiently in Sect.~\ref{sect4}.

\section{Time dependent Hartree-Fock and the boosted baryon}\label{sect3}

Let $S(x),P(x)$ denote the self-consistent scalar and pseudoscalar potentials for the baryon in the rest frame. The Dirac-TDHF equation
in a frame where the baryon is moving with velocity $v$ reads 
\begin{equation}
\left[ {\rm i} \gamma^{\mu}\partial_{\mu} -S( \gamma (x-vt)) -{\rm i}\gamma_5 \lambda P( \gamma (x-vt))     \right] \psi_{\alpha}(x,t)=0
\label{F1}
\end{equation}
with $\gamma=(1-v^2)^{-1/2}$ (not to be confused with the confinement parameter) and the self-consistency conditions
\begin{eqnarray}
S(\gamma (x-vt)) & = &  - g^2 \sum_{\alpha}^{\rm occ} \bar{\psi}_{\alpha}(x,t)\psi_{\alpha}(x,t) +m_0,
\nonumber \\
P(\gamma (x-vt)) &  = & - g^2 \sum_{\alpha}^{\rm occ} \bar{\psi}_{\alpha}(x,t){\rm i}\gamma_5 \psi_{\alpha}(x,t). 
\label{F2}
\end{eqnarray}
It is straightforward to solve Eqs.~(\ref{F1},\ref{F2}) by a Lorentz boost from the rest frame of the baryon
to the frame in which it has velocity $v$. Starting point is the ansatz
\begin{equation}
\psi_{\alpha}(x,t) = {\cal N}_{\alpha}{\rm e}^{\xi \gamma_5/2} {\rm e}^{-{\rm i}E_{\alpha}t'}\phi_{\alpha}(x')
\label{F3}
\end{equation}
where $\xi= {\rm artanh}\, v$ is the rapidity parametrizing the boost 
\begin{equation}
\left( \begin{array}{c} t' \\ x' \end{array} \right) = \left( \begin{array}{rr} \cosh \xi & - \sinh \xi \\ - \sinh \xi & \cosh \xi 
\end{array} \right)
\left( \begin{array}{c} t \\ x \end{array} \right),
\label{F4}
\end{equation}
and $E_{\alpha},\phi_{\alpha}$ denote the eigenvalues and eigenspinors of the HF Hamiltonian in the rest frame.
${\cal N}_{\alpha}$ is a (real) normalization factor. Insert Eq.~(\ref{F3}) into Eq.~(\ref{F1}),
\begin{equation}
\left[ {\rm i} \gamma^0 \partial_t + {\rm i}\gamma^1 \partial_x - S(x') -{\rm i}\gamma_5 \lambda P(x') \right] {\rm e}^{\xi \gamma_5/2} {\rm e}^{-{\rm i}
E_{\alpha}t'}\phi_{\alpha}(x')=0
\label{F5}
\end{equation} 
Next we pull the spinor boost matrix $\exp (\xi \gamma_5/2)$ through the $\gamma^{\mu}$ matrices to the left and divide it out,
using
\begin{equation}
{\rm e}^{- \gamma_5 \xi/2}\left( \begin{array}{c} \gamma^0 \\ \gamma^1 \end{array} \right) {\rm e}^{ \gamma_5 \xi/2}=
\left( \begin{array}{cc} \cosh \xi & \sinh \xi \\ \sinh \xi & \cosh \xi \end{array} \right) \left( \begin{array}{c} \gamma^0
 \\ \gamma^1 \end{array} \right)
\label{F6}
\end{equation}
and 
\begin{equation}
\left( \begin{array}{c} \partial_t \\ \partial_x \end{array} \right) = \left( \begin{array}{rr} \cosh \xi & - \sinh \xi \\ - \sinh \xi & \cosh \xi 
\end{array} \right)
\left( \begin{array}{c} \partial_{t'} \\ \partial_ {x'} \end{array} \right).
\label{F7}
\end{equation}
This yields
\begin{equation}
\left[ {\rm i} \gamma^0 \partial_{t'} + {\rm i} \gamma^1 \partial_{x'} - S(x') -{\rm i} \gamma_5 \lambda P(x')\right] {\rm e}^{-{\rm i} E_{\alpha}t'}
\phi_{\alpha}(x') = 0
\label{F8}
\end{equation} 
or, equivalently, the stationary Dirac-HF equation in the rest frame,
\begin{equation}
\left[ -{\rm i}\gamma_5 \partial_x + \gamma^0 S(x) + {\rm i} \gamma^1 \lambda P(x) \right] \phi_{\alpha}(x) = E_{\alpha} \phi_{\alpha}(x).
\label{F9}
\end{equation}
The self-consistency in the boosted frame follows from the assumed self-consistency in the rest frame since
\begin{eqnarray}
\bar{\psi}_{\alpha}(x,t) \psi_{\alpha}(x,t) &=& {\cal N}_{\alpha}^2 \bar{\phi}_{\alpha}(x') \phi_{\alpha}(x'),
\nonumber \\
\bar{\psi}_{\alpha}(x,t){\rm i}\gamma_5 \psi_{\alpha}(x,t) &=& {\cal N}_{\alpha}^2 \bar{\phi}_{\alpha}(x'){\rm i}\gamma_5 \phi_{\alpha}(x').
\label{F10}
\end{eqnarray}
The normalization factor ${\cal N}_{\alpha}$ is needed so that
the sums over occupied states in Eqs.~(\ref{F2}) are done correctly in both frames of reference.

In summary, once one has solved the HF equations in the rest frame of the baryon, a standard kinematical boost is sufficient to transform the solution into a 
solution of the TDHF equations (\ref{F1},\ref{F2}). All we have to do in the following is to compute observables using the boosted
wave functions, Eq.~(\ref{F3}).

\section{Test of covariance via calculation of energy and momentum of the boosted baryon}\label{sect4}

Let us evaluate the expectation value of the operators $H$ and $P$, Eqs.~(\ref{E3}), for a moving baryon.  Eqs.~(\ref{E4},\ref{E5}) remain valid provided
we replace the rest frame single particle spinors $\phi_{\alpha}(x)$
by the boosted ones, $\psi_{\alpha}(x',t')$, and $S(x),P(x)$ by the boosted potentials $S(x'),P(x')$. We consider first
the contributions $\langle {\cal H}_{\alpha} \rangle, \langle {\cal P}_{\alpha} \rangle$ from one single particle state $\alpha$ to the sum in Eqs.~(\ref{E5}).
Using
\begin{eqnarray}
\partial_x & = & \cosh \xi \partial_{x'} - \sinh \xi \partial_{t'},
\nonumber \\
{\rm e}^{\xi \gamma_5} &=& \cosh \xi + \gamma_5 \sinh \xi,
\label{G1}
\end{eqnarray}
and eliminating the time derivative of $\psi_{\alpha}$ with the help of the Dirac-HF equation
\begin{equation}
{\rm i}\partial_{t'}\psi_{\alpha}(x',t') = \left[-{\rm i} \gamma_5 \partial_{x'}
 + \gamma^0 S(x') + {\rm i} \gamma^1 \lambda P(x')\right] \psi_{\alpha}(x',t'),
\label{G2}
\end{equation}
a large number of terms is generated which we organize (in anticipation of the result) as follows,
\begin{eqnarray}
\langle {\cal H}_{\alpha} \rangle & = & {\cal N}_{\alpha}^2 \sum_{i=1}^3 \langle {\cal H}_{\alpha}^{(i)} \rangle,
\nonumber \\
\langle {\cal P}_{\alpha} \rangle& = & {\cal N}_{\alpha}^2 \sum_{i=1}^3 \langle {\cal P}_{\alpha}^{(i)} \rangle,
\label{G3}
\end{eqnarray}
with
\begin{eqnarray}
\langle {\cal H}_{\alpha}^{(1)}\rangle & = & \cosh (2\xi)  \phi_{\alpha}^{\dagger}(x')\gamma_5 \frac{1}{\rm i}\partial_{x'}
\phi_{\alpha}(x')
\nonumber \\
& & + \frac{1}{2} \left[ m_0 + \cosh (2\xi) S(x')\right] \bar{\phi}_{\alpha}(x')\phi_{\alpha}(x')
\nonumber \\
& & + \frac{1}{2} \cosh (2\xi) \lambda P(x') \bar{\phi}_{\alpha}(x'){\rm i}\gamma_5 \phi_{\alpha}(x'),
\nonumber \\
\langle {\cal H}_{\alpha}^{(2)}\rangle & = & \sinh (2\xi)  \phi_{\alpha}^{\dagger}(x')\frac{1}{\rm i} \partial_{x'}\phi_{\alpha}(x'),
\nonumber \\
\langle {\cal H}_{\alpha}^{(3)}\rangle & = & \frac{\rm i}{2}\sinh (2\xi) \left[ S(x')\bar{\phi}_{\alpha}(x'){\rm i}\gamma_5 \phi_{\alpha}(x')\right.
\nonumber \\
& &\left. - \lambda P(x') \bar{\phi}_{\alpha}(x')\phi_{\alpha}(x')\right],
\nonumber \\
\langle {\cal P}_{\alpha}^{(1)}\rangle & = &  \sinh (2\xi) \phi_{\alpha}^{\dagger}(x')\gamma_5 \frac{1}{\rm i}\partial_{x'}
\phi_{\alpha}(x')
\nonumber \\
& & + \frac{1}{2} \sinh (2\xi) S(x')\bar{\phi}_{\alpha}(x')\phi_{\alpha}(x')
\nonumber \\
& & + \frac{1}{2} \sinh (2\xi) \lambda P(x')\bar{\phi}_{\alpha}(x'){\rm i}\gamma_5 \phi_{\alpha}(x'),
\nonumber \\
\langle {\cal P}_{\alpha}^{(2)}\rangle & = & \cosh (2 \xi)  \phi_{\alpha}^{\dagger}(x')\frac{1}{\rm i} \partial_{x'}\phi_{\alpha}(x'),
\nonumber \\
\langle {\cal P}_{\alpha}^{(3)}\rangle & = & {\rm i}\sinh^2 \xi  \left[ S(x')\bar{\phi}_{\alpha}(x'){\rm i}\gamma_5 \phi_{\alpha}(x')\right.
\nonumber \\
& &\left. - \lambda P(x') \bar{\phi}_{\alpha}(x')\phi_{\alpha}(x')\right].
\label{G4}
\end{eqnarray}
Once again we have to treat separately the localized, subtracted parts of the various densities and their constant,
asymptotic parts. Consider the localized densities first. The normalization factor ${\cal N}_{\alpha}$ is necessary to transform
the sum over continuum states from one frame into the other one.
To leading order in $1/L$ needed here, it is given by
\begin{equation}
{\cal N}_{\alpha}^2 = \frac{E}{\omega}.
\label{G5}
\end{equation}
We denote the center-of-mass (CM) frame kinematical variables by $(E,k)$, the laboratory (LAB) frame variables by $(\omega,q)$, so that
\begin{equation}
\left( \begin{array}{c} \omega \\ q \end{array} \right) = \left(\begin{array}{cc} \cosh \xi & \sinh \xi \\ \sinh \xi & \cosh \xi \end{array}\right)
\left( \begin{array}{c} E \\ k \end{array} \right).
\label{G6}
\end{equation}
When integrating over $q$, Lorentz invariance of the measure ${\rm d}q/\omega$ allows us to relate the
quantities summed over all continuum states in the LAB and CM frames. For the discrete states, ${\cal N}_{\alpha}=1$.
The 2nd issue is the integration over $x$ when going from densities to expectation values. If we work in a finite box of length $L$
and transform integration variables from $x$ to $x'$, the integration limits acquire a time dependence,
\begin{equation}
\int_{-L/2}^{L/2} {\rm d}x \to \frac{1}{\cosh \xi} \int_{-L/2 \cosh \xi - t \sinh \xi}^{L/2 \cosh \xi - t \sinh \xi} {\rm d}x'.
\label{G7}
\end{equation}
For the localized densities, we may safely extend the integration limits to $\pm \infty$ since we are eventually interested
in the limit $L\to \infty$ (for finite $L$, we assume that the times considered are such that the baryon does not yet see the walls). It is then clear that 
the $i=2$ pieces in Eq.~(\ref{G3}) do not contribute to the expectation values of $H,P$, being proportional to the baryon momentum
in the rest frame. Moreover, the $i=3$ pieces vanish due to parity (the integrand is odd under reflection).
The only non-vanishing contributions left are the $i=1$ terms. They simplify drastically even before integrating over $x'$ once we
invoke the local conservation of the energy momentum tensor, Eq.~(\ref{E20}). As a result, the contribution from the
localized densities to the expectation values of $H,P$ is simply
\begin{eqnarray}
\langle H\rangle_{\rm loc}  &=& \cosh \xi \, m_0 \int_{-\infty}^{\infty} {\rm d}x \langle \bar{\psi}\psi \rangle_{\rm loc},
\nonumber \\
\langle P\rangle_{\rm loc} &=& \sinh \xi \,  m_0 \int_{-\infty}^{\infty}{\rm d}x \langle \bar{\psi}\psi \rangle_{\rm loc}.
\label{G8}
\end{eqnarray}
Hence this part is covariant by itself, cf. the corresponding contribution to the baryon mass in Eq.~(\ref{E21}).

Now consider the constant, asymptotic terms in the densities. The expectation values for the continuum state labelled by
$k$ in the CM frame and by $q$ in the LAB frame are given by [cf. Eqs.~(\ref{E8},\ref{E11})]
\begin{eqnarray}
L \langle {\cal H}_{\alpha} \rangle_{\rm asy} & = & N \frac{2q^2+m^2+m m_0}{2 \omega}\left( 1 - \frac{1}{L} \frac{{\rm d} \delta(k(q))}{{\rm d}q}\right),
\nonumber \\
L \langle {\cal P}_{\alpha} \rangle_{\rm asy} & = & Nq \left( 1   - \frac{1}{L} \frac{{\rm d} \delta(k(q))}{{\rm d}q} \right).
\label{G9}
\end{eqnarray}
If $F$ denotes temporarily $H$ or $P$ and ${\cal F}$ the corresponding densities ${\cal H},{\cal P}$, the relation analoguous
to Eq.~(\ref{E10}) for the boosted baryon reads
\begin{eqnarray}
\langle F \rangle_{\rm asy} & = & 
 -  L \int_{-\Lambda/2}^{\Lambda/2} \frac{{\rm d}q}{2\pi}  \frac{{\rm d}}{{\rm d}q} \left[ \delta(k(q)) \langle {\cal F}_{\rm vac}(q) \rangle\right]
\nonumber \\
& = & -\frac{L}{2 \pi} \lim_{q \to \infty} \left[ \delta(k(q)) \langle  {\cal F}_{\rm vac}(q) \rangle \right.
\nonumber \\
& & \left.  - \delta(k(-q)) \langle {\cal F}_{\rm vac}(-q)\rangle \right].
\label{G10}
\end{eqnarray}
As shown above in Eq.~(\ref{E24}), for large $k$ the phase shift behaves as
\begin{equation}
\delta(k) =  \frac{\eta}{k} + {\rm O} \left( \frac{1}{k^3} \right)
\label{G11}
\end{equation}
where 
\begin{equation}
\eta = \lim_{k\to \infty} k \delta(k) = - \frac{1}{2} \int_{-\infty}^{\infty} {\rm d}x \left( S^2+\lambda P^2-m^2 \right)
\label{G11a}
\end{equation}
and hence
\begin{equation}
\delta(k(q))  \approx  {\rm e}^{\mp \xi} \frac{\eta}{q} \qquad {\rm for\ } q \to \pm \infty
\label{G12}
\end{equation}
for negative energy states (use $k=\cosh \xi \, q - \sinh \xi \, \omega$). Together with the trivial free asymptotic behavior
\begin{eqnarray}
L \langle {\cal H}_{\rm vac}(q) \rangle & \approx & -|q|N
\nonumber \\
L \langle {\cal P}_{\rm vac}(q) \rangle & \approx & qN
\label{G13}
\end{eqnarray}
for $q \to \pm \infty$, this yields the following final result for the contribution from the 
asymptotic densities to energy and momentum of the boosted baryon,
\begin{eqnarray}
\langle H \rangle_{\rm asy}   & = & \cosh \xi \, \frac{N}{\pi} \lim_{k\to \infty} k \delta(k),
\nonumber \\
\langle P \rangle_{\rm asy} & = & \sinh \xi \, \frac{N}{\pi} \lim_{k\to \infty} k \delta(k).
\label{G14}
\end{eqnarray}
Eqs.~(\ref{E21},\ref{G8},\ref{G14}) then confirm the covariance of the spectrum,
\begin{eqnarray}
\langle H \rangle & = & \cosh \xi \, M_{\rm B},
\nonumber \\
\langle P \rangle & = & \sinh \xi \, M_{\rm B},
\label{G15}
\end{eqnarray}
even without invoking the explicit solution of the GN baryon. This is important since such a solution is known analytically only for the 
GN model ($\lambda=0$), but not for the massive NJL model ($\lambda=1$).

\section{Quark and antiquark distribution functions for the boosted baryon}\label{sect5}

In the preceding sections, we have verified that the TDHF equation for a baryon moving with velocity $v$ can be solved by a Lorentz boost
of the HF solution in the rest frame. The covariant energy-momentum relation was found in such an approach. This supports strongly
that the Dirac-TDHF approach is the correct procedure in the large $N$ limit. In general, being able to work in different Lorentz frames
does not yet yield any new physics insights, even if it is satisfying from a theoretical point of view. Nevertheless, there is one interesting application
which we shall pursue in this and the following section: The study of distribution functions for quarks and antiquarks in momentum space as a
function of baryon velocity. By going to the infinite momentum frame limit, we can then extract structure functions closely related to those
appearing in QCD analyses of deep-inelatic electron-proton scattering. Previous attempts to determine structure functions
in various quark models (e.g., the MIT bag model) have suffered from a lack of covariance of the underlying formalism, a problem which we do not share.
On the other hand, structure functions are exceedingly hard to compute in lattice QCD where one is typically able to evaluate the lowest few moments 
only (see e.g. \cite{38,39}). Therefore it may be useful to have a covariant field theoretic model where one can compute structure functions exactly. 
For this purpose, we
need the explicit HF wave functions of the baryon which are known analytically only for the massless and massive GN models [$\lambda=0$ in Eq.~(\ref{A1})].
In order to get as much analytical insight as possible, we restrict ourselves to these variants of the GN model family from here on and deal exclusively 
with DHN baryons \cite{3,5,9}. 

The fermion field operator can be expanded either in terms of the free basis, $\phi_k^{(\pm)}$, or the HF basis, $\psi_n^{(\pm)}$,
\begin{eqnarray}
\psi(x) & = & \sum_k \left( a_k \phi_k^{(+)}(x)+ b_k \phi_k^{(-)}(x) \right)
\nonumber \\
& = & \sum_n \left( A_n \psi_n^{(+)}(x) + B_n \psi_n^{(-)}(x) \right).
\label{H1}
\end{eqnarray}
The superscript $(\pm)$ refers to positive and negative energy states. The annihilation operators in the two bases are related by
the Bogoliubov transformation
\begin{eqnarray}
a_k & = & \sum_n \left\{ (\phi_k^{(+)}, \psi_n^{(+)}) A_n + (\phi_k^{(+)} , \psi_n^{(-)}) B_n \right\},
\nonumber \\
b_k & = & \sum_n \left\{ (\phi_k^{(-)}, \psi_n^{(+)}) A_n + (\phi_k^{(-)} , \psi_n^{(-)}) B_n \right\}.
\label{H2}
\end{eqnarray}
We define (momentum space) distribution functions for positive and negative energy fermions as
\begin{eqnarray}
\langle {\rm HF}| a_k^{\dagger} a_k |{\rm HF} \rangle &=& \sum_{n}^{\rm occ} |(\phi_k^{(+)},\psi_n^{(+)})|^2 + 
\sum_{n}^{\rm occ} |(\phi_k^{(+)},\psi_n^{(-)})|^2,
\nonumber \\
\langle {\rm HF}| b_k^{\dagger} b_k |{\rm HF} \rangle &=& \sum_{n}^{\rm occ} |(\phi_k^{(-)},\psi_n^{(+)})|^2 +
\sum_{n}^{\rm occ} |(\phi_k^{(-)},\psi_n^{(-)})|^2,
\nonumber \\
\label{H3}
\end{eqnarray}
where $|{\rm HF}\rangle$ denotes the HF baryon state and the sums run over all occupied states, i.e., the filled
Dirac sea and the valence level. In view of the physical interpretation we convert the negative energy fermions
into antiparticles by means of a standard particle-hole conjugation,
\begin{equation}
b_k^{\dagger}b_k  = d_{-k} d_{-k}^{\dagger} = 1 - d_{-k}^{\dagger}d_{-k},
\label{H4}
\end{equation}
and identify momentum distribution functions for quarks and antiquarks as follows,
\begin{eqnarray}
W_q(k) &=& \langle {\rm HF} |a_k^{\dagger} a_k |{\rm HF} \rangle,
\nonumber \\
W_{\bar{q}}(k) & = & 1 - \langle {\rm HF}| b_{-k}^{\dagger} b_{-k} |{\rm HF} \rangle.
\label{H5}
\end{eqnarray}
Note that there is no summation over flavor, so that all momentum distributions refer to a single flavor.

For DHN baryons and antibaryons (fermion number $\pm \nu N$), we display the discrete and continuum states
explicitly,
\begin{eqnarray}
W_q^B(k) & = & \nu |(\phi_k^{(+)},\psi_0^{(+)})|^2 + |(\phi_{k}^{(+)},\psi_0^{(-)})|^2 
\nonumber \\
& & + \sum_n^{\rm cont}  |(\phi_{k}^{(+)},\psi_n^{(-)})|^2,
\nonumber \\
W_{\bar{q}}^B(k) & = & 1 - \nu |(\phi_{-k}^{(-)},\psi_0^{(+)})|^2 - |(\phi_{-k}^{(-)},\psi_0^{(-)})|^2 
\nonumber \\
& & - \sum_n^{\rm cont}  |(\phi_{-k}^{(-)},\psi_n^{(-)})|^2,
\nonumber \\
W_q^{\bar{B}}(k) & = & (1-\nu) |(\phi_{k}^{(+)},\psi_0^{(-)})|^2 + \sum_n^{\rm cont}  |(\phi_{k}^{(+)},\psi_n^{(-)})|^2,
\nonumber \\
W_{\bar{q}}^{\bar{B}}(k) & = & 1 - (1-\nu) |(\phi_{-k}^{(-)},\psi_0^{(-)})|^2 - \sum_n^{\rm cont}  |(\phi_{-k}^{(-)},\psi_n^{(-)})|^2.
\nonumber \\
\label{H6}
\end{eqnarray}
Remember that the GN model is charge conjugation symmetric, so that all single particle states come in pairs 
with opposite energy. Eqs.~(\ref{H6}) merely reflect the different ways in which the discrete levels are
filled. In the baryon, the negative energy state is completely filled with $N$ fermions, the positive energy state partially with filling fraction 
$\nu=n/N$. In the antibaryon, the positive energy state is empty whereas the negative energy state is filled with fraction
$1-\nu$.
By charge conjugation ($C$), we must have
\begin{equation}
W_q^{B}(k)=W_{\bar{q}}^{\bar{B}}(k), \qquad W_{\bar q}^{B}(k)=W_{q}^{\bar{B}}(k).
\label{H7}
\end{equation}
This enables us to express the $C$-odd combination in terms of valence quantities only and to simplify somewhat the $C$-even
combination,
\begin{eqnarray}
W_q^B(k)-W_{\bar{q}}^B(k) & = & \nu \left\{ |(\phi_k^{(+)},\psi_0^{(+)})|^2+ |(\phi_k^{(+)},\psi_0^{(-)})|^2 \right\},
\nonumber \\
W_q^B(k) + W_{\bar{q}}^B(k) & = & \nu |(\phi_k^{(+)},\psi_0^{(+)})|^2
\nonumber \\
& &  + (2-\nu)  |(\phi_k^{(+)},\psi_0^{(-)})|^2
\nonumber \\
& & + 2 \sum_n^{\rm cont} |(\phi_k^{(+)},\psi_n^{(-)})|^2.
\label{H8}
\end{eqnarray}
Since the single particle wave functions for the DHN baryon are known,
the distribution functions for the discrete states can easily be evaluated analytically.
We refer the reader to \cite{5} for the relevant detailed wave functions in the rest frame of the 
baryon. Upon boosting these spinors according to Eq.~(\ref{F3}) and using units such that the dynamical fermion mass in the
vacuum has the value $m=1$ from now on, we find
\begin{eqnarray}
\frac{L}{2\pi} |(\phi_k^{(+)},\psi_0^{(+)})|^2 & = & \frac{\pi \left( \cos(\alpha \Delta_-) - q \sin (\alpha \Delta_-)+E_q\right)}
{8 E_k y  \cosh^2 \xi \cosh^2 (\beta \Delta_-) }, 
\nonumber \\
\frac{L}{2\pi} |(\phi_k^{(+)},\psi_0^{(-)})|^2 & = & -\frac{\pi \left( \cos(\alpha \Delta_+) + q \sin (\alpha \Delta_+)-E_q \right)}
{8 E_k y  \cosh^2 \xi \cosh^2 (\beta \Delta_+) }, 
\nonumber \\
& & \ 
\label{H9}
\end{eqnarray}
with
\begin{eqnarray}
\alpha & = & \frac{2 c_0}{y \cosh \xi}, \quad c_0 \ = \ \frac{1}{2} {\rm artanh}\, y
\nonumber \\
\beta & = & \frac{\pi}{2 y \cosh \xi},
\nonumber  \\
\Delta_{\pm} & = & \sqrt{1-y^2} \sinh \xi \pm k,
\label{H10}
\end{eqnarray}
and the boosted kinematical variables
\begin{equation}
\left( \begin{array}{c} E_q \\ q \end{array} \right) = \left(\begin{array}{cc} \cosh \xi & \sinh \xi \\ \sinh \xi & \cosh \xi \end{array}\right)
\left( \begin{array}{c} E_k \\ k \end{array} \right).
\label{H11}
\end{equation}
The factors $L/(2\pi)$ in Eq.~(62) have been introduced with regard to the limit $L\to \infty.$
The parameter $y$ is determined by the occupation fraction $\nu$ of the positive energy
valence state of the DHN baryon \cite{3},
\begin{equation}
y = \sin \left( \frac{\pi\nu}{2} \right).
\label{H11a}
\end{equation}
Note also the symmetry relation which can be used to obtain other related matrix elements,
\begin{equation}
|(\phi_k^{(\sigma)},\psi_0^{(\eta)})|^2 = |(\phi_{-k}^{(-\sigma)},\psi_0^{(-\eta)})|^2 \qquad (\sigma,\eta  =\pm 1).
\label{H12}
\end{equation}
Matrix elements involving continuum HF states have been computed as follows: We start from box normalized spinors
and boost the HF spinors \cite{5} from the rest frame to the moving frame. The relevant matrix element is
\begin{equation}
Z_{\rm cont} = \left( \frac{L}{2\pi} \right)^2| ( \phi_k^{(+)},\psi_K^{(-)})|^2
\label{H16}
\end{equation} 
with the analytical result 
\begin{equation}
Z_{\rm cont} = \frac{(K-q)\sin (\alpha C) - (qK+1)\cos(\alpha C) + E_q E_K}{ 8 \cosh^2 \xi (K^2+y^2)E_k E_Q \sinh^2(\beta C)}.
\label{H17}
\end{equation}
Here, 
\begin{equation}
\left( \begin{array}{c} E_Q \\ Q \end{array} \right) = \left(\begin{array}{cc} \cosh \xi & - \sinh \xi \\ - \sinh \xi & \cosh \xi \end{array}\right)
\left( \begin{array}{c} E_K \\ K \end{array} \right),
\label{H17a}
\end{equation}
$\alpha$ and $\beta$ are defined as in Eq.~(\ref{H10}), and
\begin{equation}
C  =  k-Q.
\label{H18}
\end{equation}
In the course of this calculation, one runs into the Fourier transform of $\tanh z$ with finite integration limits. This was done
as follows: Add and subtract the function ${\rm sgn} z$. In the Fourier transform of $(\tanh z-{\rm sgn}\, z)$, we may safely extend the 
integration limits to $\pm \infty$ in the large $L$ limit, 
\begin{equation}
\int_{-\infty}^{\infty} {\rm d}z (\tanh z - {\rm sgn}\, z ) {\rm e}^{-{\rm i}pz}= -2 {\rm i} \left( \frac{\pi /2}{\sinh (\pi p/2)}- \frac{1}{p} \right).
\label{H19}
\end{equation} 
The integral over ${\rm sgn}\, z$ is then evaluated with finite integration limits. 
Performing the limit $L\to \infty$ at the end, all momenta go over into continuous variables and we normalize the momentum distributions
accordingly. Summarizing, our final result for the discrete and continuum contributions is given in closed analytical form up to a one-dimensional integration,
\begin{eqnarray}
W_q^B(k) & = & \nu W_1(k) + W_2(k) + W_3(k),
\nonumber \\
W_{\bar{q}}^B(k) & = &  (1-\nu) W_2(k) + W_3(k),
\nonumber \\
W_{\rm val}^B(k) & = & \nu W_1(k),
\label{H20}
\end{eqnarray}
with 
\begin{eqnarray}
W_1(k) & = & \frac{\pi \left( \cos(\alpha \Delta_-) - q \sin (\alpha \Delta_-)+E_q\right)}
{8 E_k y  \cosh^2 \xi \cosh^2 (\beta \Delta_-) }, 
\nonumber \\
W_2(k) & = & -\frac{\pi \left( \cos(\alpha \Delta_+) + q \sin (\alpha \Delta_+)-E_q \right)}
{8 E_k y  \cosh^2 \xi \cosh^2 (\beta \Delta_+) }, 
\nonumber \\
W_3(k) & = & 
\nonumber \\
\int_{-\infty}^{\infty} & {\rm d}Q & \frac{(K-q)\sin (\alpha C) - (qK+1)\cos(\alpha C) + E_q E_K}{ 8 \cosh^2 \xi (K^2+y^2)E_k E_Q \sinh^2(\beta C)}.
\nonumber \\
\label{H21}
\end{eqnarray}
These distribution functions are normalized according to the following ``sum rules" to baryon number and baryon momentum,
\begin{eqnarray}
\nu \ = \ \frac{n}{N} & = & \int_{- \infty}^{\infty} {\rm d}k \left( W_q^B(k)-W_{\bar{q}}^B(k) \right),
\nonumber \\
\frac{P_B}{N} & = & \int_{-\infty}^{\infty} {\rm d}k k  \left( W_q^B(k) + W_{\bar{q}}^B(k) \right).
\label{H22}
\end{eqnarray}
\begin{figure}
\begin{center}
\epsfig{file=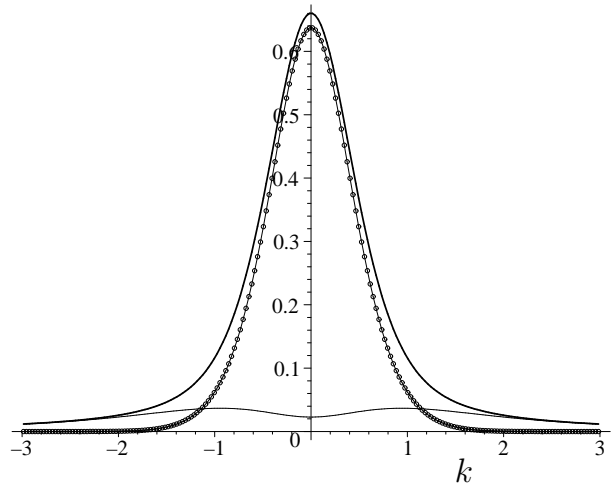,height=8cm,width=6.4cm,angle=270}
\caption{Quark ($W_q^B$, thick curve), antiquark ($W_{\bar{q}}^B$, thin curve) and valence quark ($W_{\rm val}^B$, dotted curve) distribution
functions for baryon at rest versus fermion momentum. The parameters are $\nu=0.75, \gamma=0$, units $m=1$.}
\label{fig1}
\end{center}
\end{figure}

\begin{figure}
\begin{center}
\epsfig{file=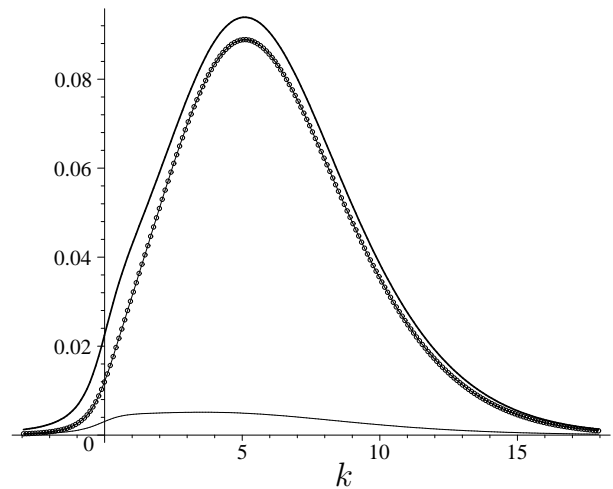,height=8cm,width=6.4cm,angle=270}
\caption{Same as Fig.~\ref{fig1}, but for a boosted baryon with momentum $P_B/N=5$. Note the changed scale on both axes.}
\label{fig2}
\end{center}
\end{figure}

\begin{figure}
\begin{center}
\epsfig{file=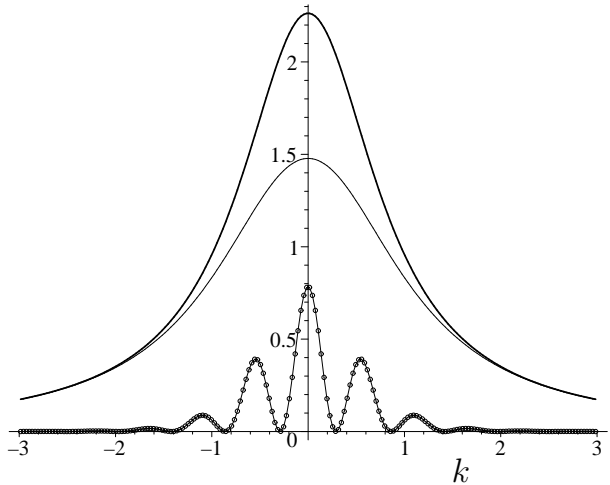,height=8cm,width=6.4cm,angle=270}
\caption{Same as Fig.~\ref{fig1}, baryon at rest but different occupation fraction. Parameters $\nu=0.99999,\gamma=0$.}
\label{fig3}
\end{center}
\end{figure}

\begin{figure}
\begin{center}
\epsfig{file=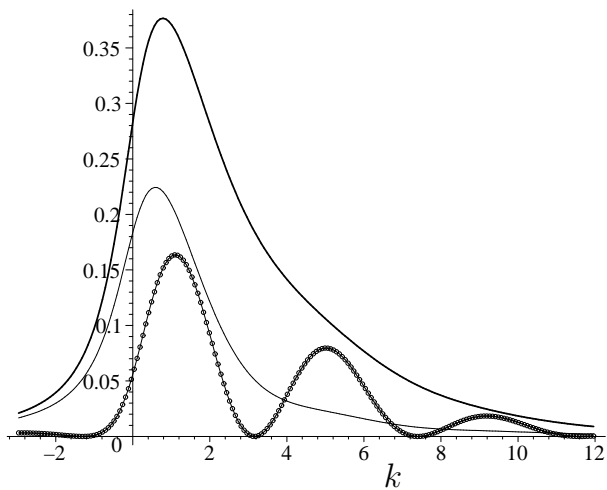,height=8cm,width=6.4cm,angle=270}
\caption{Same as Fig.~\ref{fig3}, but for boosted baryon with momentum $P_B/N=5$ and adjusted scale.}
\label{fig4}
\end{center}
\end{figure}

Let us illustrate the behavior of the distribution functions with a few examples for DHN baryons in the massless GN model. We first choose
a moderate occupation fraction of the valence level ($\nu=0.75$). In Fig.~{\ref{fig1}, we show the quark and antiquark distribution functions $W_q^B$
and $W_{\bar{q}}^B$ for a baryon at rest, together with the contribution from the positive energy discrete state, $W_{\rm val}^B$. 
Fig.~\ref{fig2} is the corresponding plot for baryon momentum $P_B/N=5$ (in units where $m=1$). At this value of $\nu$,
the contributions from the Dirac sea and from antiquarks are rather small, with little dependence on baryon momentum. 
This picture changes if we go into the regime where the bound state is highly relativistic by choosing $\nu=0.99999$. As shown in Figs.~\ref{fig3}
and \ref{fig4}, here the valence quark distribution function bears little resemblance with the full quark result, showing an enhanced role
of the Dirac sea. To quantify the dependence of sea effects on total baryon momentum, we have also integrated the quark and 
antiquark distribution  functions over all momenta $k$. 
At this point it is convenient to introduce hatted quantities for baryon momentum, baryon mass and fermion number
\begin{equation}
\hat{P}_B = \frac{P_B}{N}, \qquad \hat{M}_B = \frac{M_B}{N}, \qquad \hat{N}_f = \frac{N_f}{N}
\label{H22a}
\end{equation}
to get rid of trivial $N$-dependences.
In Figs.~\ref{fig5} and \ref{fig6} are shown the contributions from quarks and antiquarks to  
reduced fermion number versus reduced baryon momentum. At $\nu=0.75$ (Fig.~\ref{fig5}), sea effects are small
everywhere, quickly reaching some (non-zero) asymptotic value. At $\nu=0.99999$ (Fig.~\ref{fig6}), they drop rapidly between $\hat{P}_B=0$ and $\hat{P}_B=10$,
but then stay constant at a sizeable level. The lesson we draw from this is that antiquark effects apparently do not disappear in the infinite
momentum frame,  
although they are reduced significantly as compared to the baryon at rest. This is in contrast to the earlier observation that antiparticle effects in the structure of
mesons are completely quenched in the infinite momentum frame \cite{40}. It suggests that a light-cone approach to baryons may be less efficient
than for mesons, at least in the large $N$ limit.
\begin{figure}
\begin{center}
\epsfig{file=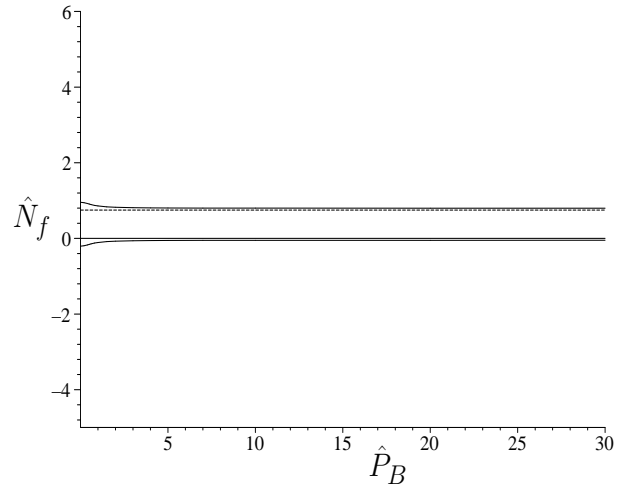,height=8cm,width=6.4cm,angle=270}
\caption{Dashed straight line: reduced fermion number $\hat{N}_f=\nu=0.75$ of DHN baryon. The solid curves show how fermion number is 
made up from quarks (upper curve, positive contribution) and antiquarks (lower curve, negative contribution), as a function of baryon
momentum. The asymptotic values reached around $\hat{P}_B=5$ are 0.80 quarks and 0.05 antiquarks, as compared to 0.95 quarks and
0.20 antiquarks at $\hat{P}_B=0$.} 
\label{fig5}
\end{center}
\end{figure}

\begin{figure}
\begin{center}
\epsfig{file=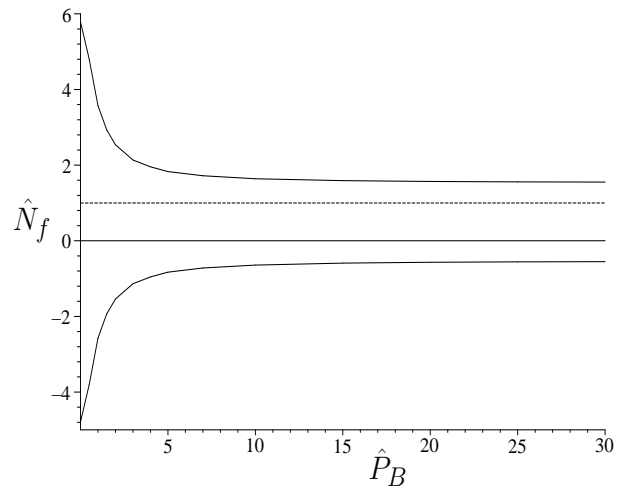,height=8cm,width=6.4cm,angle=270}
\caption{Same as Fig.~\ref{fig5}, but different fermion number $\hat{N}_f=\nu=0.99999$. Here the asymptotic values reached
around $\hat{P}_B=10$ are 1.55 quarks and 0.55 antiquarks, as compared to 5.79 quarks and 4.79 antiquarks at $\hat{P}_B=0$.}
\label{fig6}
\end{center}
\end{figure}

\section{Infinite momentum frame and structure functions}\label{sect6}

To facilitate the infinite momentum frame limit, we scale the momentum variable with the reduced baryon momentum,
\begin{equation}
k = x \hat{P}_B,
\label{H23}
\end{equation}
and introduce the rescaled densities
\begin{equation}
w_q^B(x) =  \hat{P}_B W_q^B(x {\hat P}_B), \quad w_{\bar{q}}^B(x) = \hat{P}_B W_{\bar{q}}^B(x {\hat P}_B).
\label{H24}
\end{equation}
In this form, the limit $\xi \to \infty$ can readily be taken. Note that since $\hat{P}_B$ is defined without the factor $N$, $x$ is not
restricted to $[0,1]$ in the infinite momentum frame like the standard Bjorken variable, but rather to $[0,N]$ 
(i.e., the positive half-axis in the limit $N\to \infty$). The result for the positive and negative energy discrete states is
\begin{equation}
w_{\rm disc}^{(\pm)}(x) =  \frac{\pi {\hat M}_B \left( 1-\sin \kappa_{\mp} \right)}{4y \cosh^2 (\pi \kappa_{\mp}/(4c_0))}
\label{H25}
\end{equation}
with 
\begin{equation}
\kappa_{\pm} = \frac{2 c_0 \left( \sqrt{1-y^2} \pm \hat{M}_B x\right)}{y}.
\label{H26}
\end{equation}
The negative energy continuum contribution reduces to
\begin{equation}
w_{\rm cont}^{(-)}(x) = \int_0^{\infty} {\rm d} q F(q,x)
\label{H27}
\end{equation}
with 
\begin{equation}
F(q,x) = \frac{{\hat M}_B \left[ (q^2 -1) \cos \kappa  -2 q \sin  \kappa +q^2 +1 \right]}{2  \left[ (q^2 -1)^2+4 y^2q^2\right]
\sinh^2 (\pi \kappa/(4c_0))}
\label{H28}
\end{equation}
and 
\begin{equation}
\kappa = \frac{2 c_0 (\hat{M}_B x+q)}{y}.
\label{H29}
\end{equation}
From Eqs.~(\ref{H25}--\ref{H29}), full quark and antiquark structure functions and the valence quark structure function for the baryon can be obtained as follows,
\begin{eqnarray}
w_q^B(x) & = & \nu w_{\rm disc}^{(+)}(x) + w_{\rm disc}^{(-)}(x) + w_{\rm cont}^{(-)}(x),
\nonumber \\
w_{\bar q}^B(x) & = & (1-\nu) w_{\rm disc}^{(-)}(x) + w_{\rm cont}^{(-)}(x),
\nonumber \\
w_{\rm val}^B(x) & = &  \nu w_{\rm disc}^{(+)}(x).
\label{H30}
\end{eqnarray}
They are normalized according to 
\begin{eqnarray}
\nu & = & \int_0^{\infty} {\rm d}x \left( w_q^B(x)-w_{\bar{q}}^B (x) \right), 
\nonumber \\
1 & = & \int_0^{\infty} {\rm d}x \, x \left( w_q^B(x)+w_{\bar{q}}^B (x) \right). 
\label{H31}
\end{eqnarray}
Watch out the integration limits characteristic for the large $N$ limit.
Finally we recall the relations between $\nu, y$ and ${\hat M}_B$ holding in the massive Gross-Neveu model \cite{8,9},
\begin{eqnarray}
\nu & = & \frac{2}{\pi} \left( \theta + \gamma \tan \theta \right),
\nonumber \\
y & = & \sin \theta,
\nonumber \\
{\hat M}_B & = & \frac{2}{\pi} \left( y+ \gamma \,{\rm artanh}\, y \right).
\label{H32}
\end{eqnarray}
\begin{figure}
\begin{center}
\epsfig{file=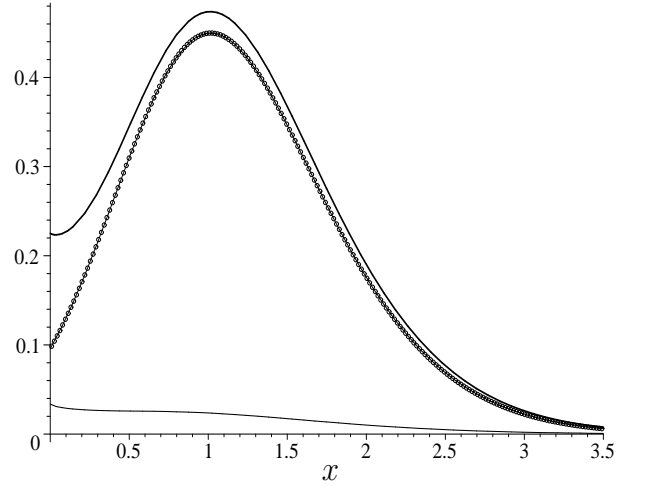,height=8cm,width=6.4cm,angle=270}
\caption{Quark ($w_q^B$, thick curve), antiquark ($w_{\bar{q}}^B$, thin curve) and valence quark ($w_{\rm val}^B$, dotted curve) structure
functions for baryon versus rescaled fermion momentum. The parameters are $\nu=0.75, \gamma=0$. This graph can be regarded as the
infinite momentum frame limit of Figs.~\ref{fig1},\ref{fig2} with appropriately rescaled axes. All structure functions drop to zero at $x=0$
and vanish identically for $x<0$.}
\label{fig7}
\end{center}
\end{figure}
\begin{figure}
\begin{center}
\epsfig{file=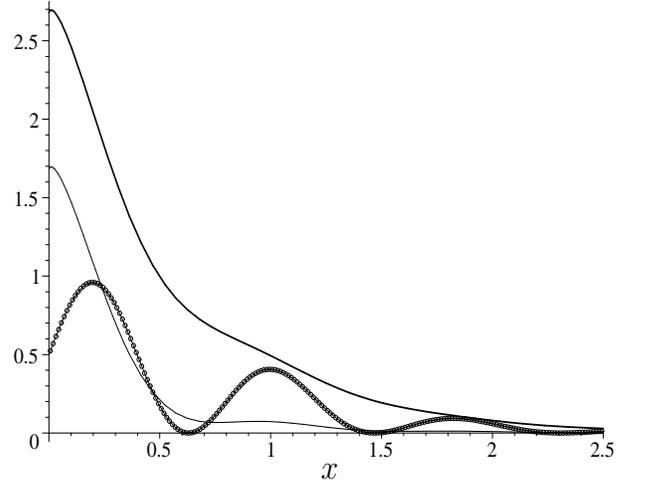,height=8cm,width=6.4cm,angle=270}
\caption{Same as Fig.~\ref{fig7} for different occupation fraction $\nu=0.99999$. Continuation of Figs.~\ref{fig3},\ref{fig4} to the 
infinite momentum frame.}
\label{fig8}
\end{center}
\end{figure}
\begin{figure}
\begin{center}
\epsfig{file=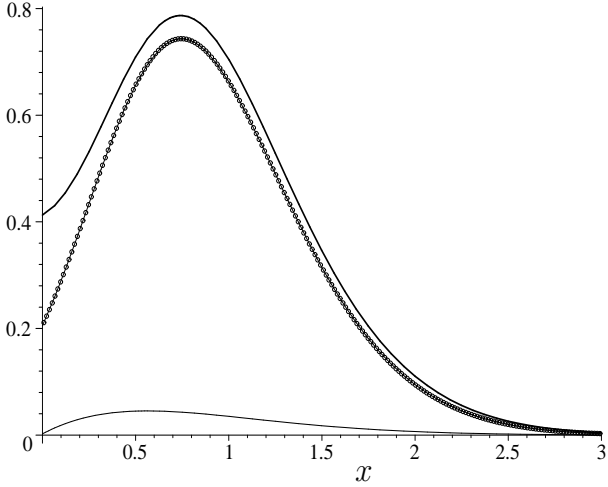,height=8cm,width=6.4cm,angle=270}
\caption{Same as Fig.~\ref{fig8} but for massive GN model with parameters $\nu=0.99999,\gamma=0.1$. Comparison with Fig.~\ref{fig7}
shows that introducing a bare quark mass has a similar effect as reducing the occupation fraction, driving the baryon into the 
non-relativistic regime.} 
\label{fig9}
\end{center}
\end{figure}
In Figs.~\ref{fig7} and {\ref{fig8}, we show by way of example structure functions for the massless GN model at $\nu=0.75$
and $\nu=0.99999$. Apart from trivial rescalings of both axes, these plots can be regarded as the continuation of Figs.~\ref{fig1}--\ref{fig4}
to the infinite momentum frame. We observe that in the infinite momentum frame sea effects are important for almost complete filling of the 
valence level. They can be reduced by choosing a smaller occupation. An alternative way to suppress sea effects is to switch on the bare
quark mass while keeping the occupation  the same, see Fig.~\ref{fig9}. For the same occupation fraction as in Fig.~\ref{fig8} and the moderate
value $\gamma=0.1$ of the confinement parameter, one gets a picture in qualitative agreement with the one at $\gamma=0$ and $\nu=0.75$
in Fig.~\ref{fig7}.
  
Finally, we illustrate the influence of the bare quark mass on sea effects with the help of integrated quantities. Here we consider full occupation ($\nu=1$)
and vary the confinement parameter $\gamma$. 
\begin{figure}
\begin{center}
\epsfig{file=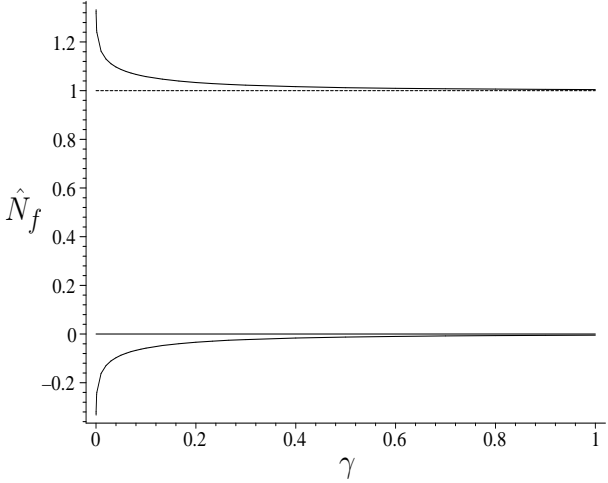,height=8cm,width=6.4cm,angle=270}
\caption{Dashed curve: reduced fermion number $\hat{N}_f=\nu=1$ of baryon in the massive DHN model. The solid curves show how
fermion number in the infinite momentum frame is made up from quarks and antiquarks, as a function of the confinement parameter (proportional to
the bare quark mass). Both curves show a logarithmic divergence in the chiral limit $\gamma\to 0$ relevant for the kink baryon, see main text.} 
\label{fig10}
\end{center}
\end{figure}
\begin{figure}
\begin{center}
\epsfig{file=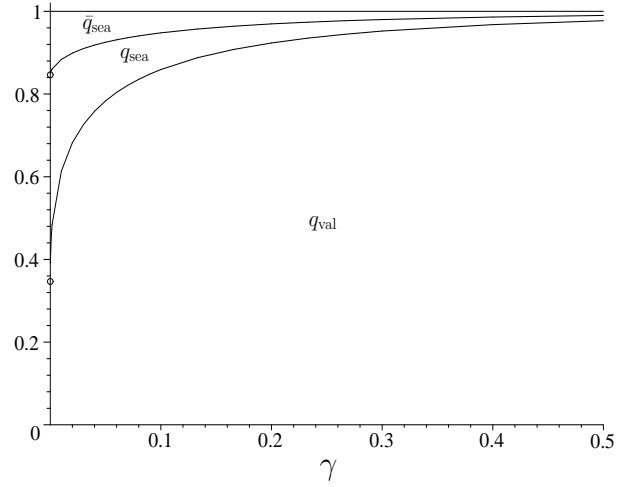,height=8cm,width=6.4cm,angle=270}
\caption{Decomposition of the total baryon momentum into valence quark, sea quark and antiquark contributions, in the infinite momentum
frame, versus $\gamma$. As in Fig.~\ref{fig10}, the valence level is fully occupied ($\nu=1$). The two circles on the 
vertical axis ($\gamma=0$) indicate the endpoints of the boundaries separating the different contributions and are 
evaluated analytically in the main text.}  
\label{fig11}
\end{center}
\end{figure}
Fig.~\ref{fig10} exhibits the individual contributions from quarks and antiquarks to total fermion
number as a way of quantifying sea effects in the infinite momentum frame. The two curves add up to the reduced fermion number 1
and exhibit strong deviations from a valence picture near $\gamma = 0$. At the point $\nu=1, \gamma=0$, the parameter $y$ tends to 1 and
the DHN baryon goes over into a kink-antikink pair at infinite separation. Our results indicate that in this limit the baryon consists of 
diverging numbers of quarks and antiquarks, even in the infinite momentum frame (below we will argue that the divergence is logarithmic).
Whereas Fig.~\ref{fig10} details the contributions from quarks and antiquarks to the total fermion number,
Fig.~\ref{fig11} decomposes the total baryon momentum in the infinite momentum frame into contributions from valence quarks,
sea quarks and antiquarks, again at $\nu=1$ as a function of $\gamma$. The boundaries between the three regions in the plot hit the
$\gamma=0$ axis at the points indicated by circles which will be determined below.

The limit $y\to 1$ is the ultrarelativistic limit for the internal baryon structure. It can be interpreted as referring to  
the kink-like baryon of the massless GN model (here with fully occupied valence level). In this limit the formulae for the 
structure functions greatly simplify,  
\begin{eqnarray}
\lim_{y\to 1}  w_q^B(x) &=&  \frac{1}{\cosh^2 x} +  \int_0^{\infty}{\rm d}q \frac{2}{(\pi^2+4q^2)\sinh^2(x+q)},
\nonumber \\
\lim_{y\to 1}  w_{\bar{q}}^B(x) &=&    \int_0^{\infty}{\rm d}q \frac{2}{(\pi^2+4q^2)\sinh^2(x+q)}.
\label{H33}
\end{eqnarray}
The $1/\cosh^2$ term is due to the discrete levels, the integral to the negative energy continuum. 
The sum rules (\ref{H31}) for $\nu=1$ are satisfied. There is an infrared divergence at $x=0$ in the continuum contribution,
\begin{equation}
\lim_{y \to 1} w_q^B(x) \approx \lim_{y \to 1} w_{\bar{q}}^B(x) \approx \frac{2}{\pi^2 x} \qquad (x \to 0).
\label{H34}
\end{equation}
It gives rise to a logarithmic divergence in the number of quarks and antiquarks upon integration over $x$, explaining the steep
rise of the curves in Fig.~\ref{fig10} towards $\gamma=0$.
If we plot the momentum distribution $x w(x)$, everything is well-behaved and we obtain Fig.~\ref{fig12} which also shows the valence
level contribution. These are the structure functions of the kink baryon. By integrating Eqs.~(\ref{H33}) over $x$ we can 
evaluate analytically the contribution to the total momentum from valence quarks, sea quarks and antiquarks.
In this way we find $\frac{1}{2}\ln 2$ for valence quarks, $\frac{1}{2}$ for sea quarks and $\frac{1}{2}-\frac{1}{2}\ln 2$ for antiquarks. 
In other words, valence quarks carry 35$\%$, sea quarks 50$\%$ and antiquarks 15$\%$ of the total (kink) baryon momentum in the 
infinite momentum frame. This explains the
location of the points on the $\gamma=0$ axes drawn in Fig.~\ref{fig11}.
\begin{figure}
\begin{center}
\epsfig{file=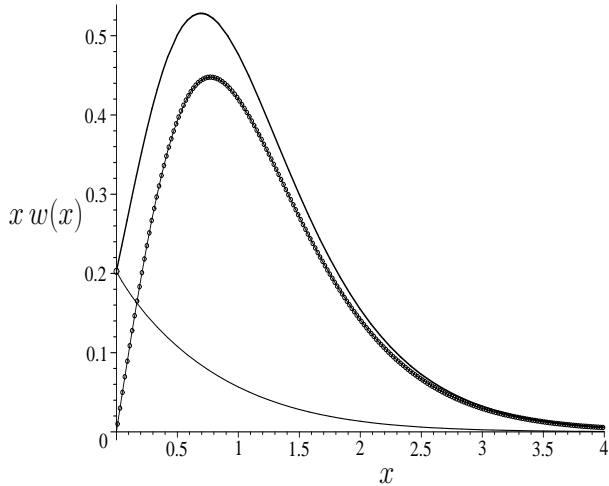,height=8cm,width=6.4cm,angle=270}
\caption{Momentum distribution functions $x w(x)$ in the infinite momentum frame and in the kink limit $y=1$.  
Quark ($x w_q^B(x)$, thick line), antiquark ($xw_{\bar{q}}^B(x)$, thin line) and valence quark ($xw_{\rm val}^B(x)$, dotted line) structure
functions are exhibited. The solid curves hit the $x=0$ axes at the point $2/\pi^2$ in accordance with Eq.~(\ref{H34}).}  
\label{fig12}
\end{center}
\end{figure}

\section{Summary and outlook}\label{sect7}

In the past, exactly solvable QFT models have given insights mostly into static problems: dynamical masses of constituents,
the structure and mass of bound states, phase diagrams at finite temperature and chemical potential. All of this touches upon non-perturbative issues
of interest to strong interaction physics. A particularly rewarding class of such toy models are large $N$ fermionic models in 1+1 dimensions
of Gross-Neveu type. In the present work, a first attempt was made to widen the scope of these studies towards time-dependent  questions. Due to the fact
that lattice gauge theories can only be solved in Euclidean time, dynamical non-perturbative phenomena are even more elusive in
realistic theories than static ones, although they may be of considerable interest from the physics point of view. 

At first, we wanted to verify that the large $N$ limit preserves covariance. Although no problems are expected on general grounds, we were
interested in a detailed study to see whether one can actually boost a composite, relativistic bound state in practice.
We found that by generalizing Dirac-HF to its time-dependent version, Dirac-TDHF, it is indeed possible to boost baryons to any frame
and to confirm the covariant relation between energy and momentum. This underlines the advantage of using QFT
toy models rather than more phenomenological quark models, where covariance is always violated at some level. We have learned how 
covariance is restored at the end of a calculation involving a non-covariant momentum cutoff and a finite box at intermediate
stages. Renormalizability is a decisive feature here.
As a by-product, we have understood cancellations observed in former calculations of the baryon mass in the rest frame and derived a simple
formula relating the mass with the HF potentials.
  
As is well known, internal motion and CM motion do not decouple in relativistic bound states, unlike in non-relativistic quantum mechanics.
In the GN model, one can
control the extent to which the internal motion is relativistic by means of the parameters $\nu$ (occupation fraction of the valence level)
and $\gamma$ (related to the bare mass). Through the choice of velocity $v$ on the other hand we can choose the degree to which the CM
motion is relativistic. Exploring the full range of these parameters should enable one to better understand how internal and overall motions are
interwoven.   

One way of analyzing how the structure of the bound state depends on its velocity is through momentum distribution of quarks and antiquarks.
Since all the wave functions are known analytically in the case of the GN model with discrete chiral symmetry,
such a calculation can be done exactly and requires only a one-dimensional numerical integration. 
We were interested in the evolution of momentum distributions from the rest frame (where one has some physical intuition from atoms or nuclei)
to the infinite momentum frame, which is accessible in nature through deep inelastic scattering but where one's intuition is in general less developed. 
We illustrated our results with a few examples where relativistic effects were either weak or strong, as evidenced
by the size of sea effects. We had actually hoped to be able to exploit covariance in order to simplify the HF problem, similarly to what
has already been achieved for the meson Bethe-Salpeter equation. Unfortunately, we did not see decoupling of the Dirac sea  in the baryon case,
although the sea effects are reduced in the infinite momentum frame. This sheds light on the possible use of light-cone quantization for baryons.

By performing the limit to the infinite momentum frame, the momentum distributions go over into functions closely related to structure functions
in real QCD. They satisfy exact sum rules, but the variable similar to Bjorken $x$ cannot be restricted to the interval [0,1] but only the positive half 
axis in the large $N$ limit. Of particular interest is the kink limit where the formulae simplify further. The ultrarelativistic character of the bound state
gives rise to a diverging number of quarks and antiquarks (coming from the low  $x$ region) which has little in common with the non-relativistic,
valence-level type baryon. 

Encouraged by these first results, we plan to address more demanding dynamical problems like kink-kink scattering, acceleration of baryons
by external fields or non-equilibrium thermodynamics in the future. This may even be of some interest for the parallel world of condensed matter
physics where the DHN baryons live a life as solitons, polarons and excitons in quasi-onedimensional systems, important for example for 
conduction properties of doped polymers \cite{36}.

\section*{Acknowledgement}

We should like to thank Falk Bruckmann and Matthias Burkardt for a helpful correspondence. 
This work has been supported in part by the Studienstiftung des deutschen Volkes (W.B.) and 
by the DFG under grant TH 842/1-1.

\end{document}